\documentclass[%
 reprint,
 amsmath,amssymb,
 prl, floatfix, showpacs
]{revtex4-1}

\usepackage{cases}
\usepackage{dcolumn}
\usepackage{bm}
\usepackage{amssymb}
\usepackage{tikz}
\usepackage{amsmath}
\usepackage{mathtools}
\usepackage{verbatim}
\usepackage[english]{babel}

\makeatletter
\adddialect\l@English\l@english
\makeatother

\begin{document}

\title{Beam engineering for selective and enhanced coupling to multipolar resonances}

\author{Tanya Das}
\author{Prasad P. Iyer}
\author{Jon A. Schuller}
 \email{jonschuller@ece.ucsb.edu}
\affiliation{%
Department of Electrical and Computer Engineering, University of California, Santa Barbara, \\ Santa Barbara, California 93106
}%

\date{\today}

\begin{abstract}
Multipolar electromagnetic phenomena in sub-wavelength resonators are at the heart of metamaterial science and technology. In this letter, we demonstrate selective and enhanced coupling to specific multipole resonances via beam engineering. We first derive an analytical method for determining the scattering and absorption of spherical nanoparticles (NPs) that depends only on the local electromagnetic field quantities within an inhomogeneous beam. Using this analytical technique, we demonstrate the ability to drastically manipulate the scattering properties of a spherical NP by varying illumination properties and demonstrate the excitation of a longitudinal quadrupole mode that cannot be accessed with conventional illumination. This work enhances the understanding of fundamental light-matter interactions in metamaterials, and lays the foundation for researchers to identify, quantify, and manipulate multipolar light-matter interactions through optical beam engineering.


\end{abstract}

\maketitle

Metamaterials derive their unique optical properties from engineered multipolar resonances in sub-wavelength structures. Understanding and quantifying the behavior of these multipolar resonances is essential for a variety of metamaterial applications, including optical antennas\cite{coenen2014directional, hancu_multipolar_2014, iyer_reconfigurable_2015}, absorptive layers for solar cells \cite{catchpole_plasmonic_2008,nakayama_plasmonic_2008,mann_dielectric_2011}, cloaking devices \cite{schurig_metamaterial_2006,alu_cloaking_2007}, superlenses \cite{fang_subdiffraction-limited_2005,kawata_plasmonics_2009}, and biosensors \cite{fard_label-free_2013}. Typically, researchers engineer multipolar light-matter interactions by modifying the size, shape, and composition of the resonators\cite{grahn_electric_2012,zhang_superlenses_2008,menzel_spectral_2014}. Here, we instead demonstrate engineering of multipolar light-matter interactions by modifying properties of the illuminating radiation. In this approach we exploit the incident illumination beam to gain information about our multipolar system.

In this work, we rigorously derive a simplified, analytical method directly from Mie theory to quantify multipolar light absorption and scattering for a spherical nanoparticle (NP) illuminated by any light source of interest.  First, we prove that dipolar and quadrupolar interactions depend only on local fields or field gradients respectively. Building from this local-field approach we investigate scattering of NPs in linearly, azimuthally, and radially polarized focused light beams. We demonstrate selective suppression and enhancement of individual multipolar modes by manipulating beam symmetries and numerical apertures. These calculations reveal a longitudinal quadrupole mode, which is completely inaccessible by conventional linearly polarized light. Additionally, we achieve selective excitation of individual multipolar modes. This work demonstrates a method for quantifying multipolar interactions in sub-wavelength particles and establishes beam engineering as a powerful method for manipulating multipolar phenomena.

Conventionally, scattering and absorption of spherical NPs is calculated with Generalized Lorenz Mie theory (GLMT) \cite{gouesbet2011generalized}, which involves expressing an incident beam as a plane wave or spherical wave expansion \cite{colak_scattering_1979}. These expansions require knowing the electric field everywhere on a planar $[\textbf{E} \left(x,y,z=z_{0} \right)]$ or spherical $[\textbf{E} \left(\theta, \phi, r=r_{0} \right)]$ surface respectively. This approach is complete, and can be used to describe interactions with spherical NPs of any size or composition. However, NPs used in plasmonics or metamaterials are typically in a size regime where only dipolar and quadrupolar modes contribute to the optical response. For these cases, we derive a greatly simplified approach for calculating scattering and absorption in inhomogeneous fields, inspired by the multipolar interaction Hamiltonian \cite{alu_guided_2009,alu_polarizabilities_2005,sersic_magnetoelectric_2011,bernal_arango_underpinning_2014,muhlig2011multipole,novotny_principles_2012}, given as,

\begin{equation}\label{eqn:Hint}
   \mathclap{H = \underbrace{-\textbf{p} \cdot \textbf{E}(\textbf{r}_p)}_{\mbox{ED}}-\underbrace{\textbf{m} \cdot\textbf{H}(\textbf{r}_p)}_{\mbox{MD}}-\underbrace{[\tensor{\textbf{Q}}\triangledown] \cdot  \textbf{E}(\textbf{r}_p)}_{\mbox{EQ}} -\underbrace{[\tensor{\textbf{G}}\triangledown] \cdot  \textbf{H}(\textbf{r}_p)}_{\mbox{MQ}}...}
\end{equation}

\noindent The light-matter interaction energy is described by a collection of multipolar terms, corresponding to electric dipole (ED), magnetic dipole (MD), electric quadrupole (EQ), and magnetic quadrupole (MQ) interactions. Each interaction term depends on a light-independent multipole moment ($\textbf{p}$ for ED, $\textbf{m}$  for MD, etc.), as well as a matter-independent electromagnetic field quantity ($\textbf{E}$ is the electric field, and $\textbf{H}$ is the magnetic field) that only depends on the local field at any point of interest ($\textbf{r}_p$) within the inhomogeneous field distribution.

These principles are combined with Mie Theory \cite{bohren_absorption_1998} (which describes the interaction of a plane wave with a spherical particle) to rigorously derive simple local-field relations that fully describe the dipolar and quadrupolar scattering interaction of NPs in inhomogeneous fields (See Supplemental Material). The total power scattered via dipolar modes is given by,

\begin{numcases}
    {P_{sca}^{dip} = \frac{3\pi}{k\omega\mu} }
        \lvert a_1\rvert^2 \lvert \textbf{E}_{inc}\left( \textbf{r}_p \right) \rvert^2, \text{ electric}\label{eqn:PscaED}\\
        Z^2 \lvert b_1\rvert^2 \lvert \textbf{H}_{inc}\left( \textbf{r}_p \right) \rvert^2, \text{ magnetic}\label{eqn:PscaMD}
\end{numcases}

\noindent where, $Z$, $\mu$, and $k$ are respectively the impedance, permeability, and wave number of the background medium. $a_n$ and $b_n$ are Mie coefficients, determined from standard plane wave illumination. Lastly, $\lvert \textbf{E}_{inc} \left(\textbf{r}_p \right) \rvert^2$ and $\lvert \textbf{H}_{inc} \left(\textbf{r}_p \right) \rvert^2$ are the electric and magnetic field intensity at the center of the NP, in the absence of the NP. In analogy with the multipolar interaction Hamiltonian, the scattering is proportional to a field-independent ``moment'', i.e., the Mie coefficient, and a particle-independent driving term, i.e., the field intensity.

The power scattered by quadrupole modes is given by,

\begin{numcases}
    {P_{sca}^{quad} = \frac{10\pi}{k^3\omega\mu} }
        \lvert a_2\rvert^2 \sum_{i,j}\lvert {Q}_{ij}\rvert^2, \text{ electric}\label{eqn:PscaEQ}\\
        Z^2 \lvert b_2\rvert^2 \sum_{i,j}\lvert {G}_{ij}\rvert^2, \text{ magnetic}\label{eqn:PscaMQ}
\end{numcases}

\noindent where the driving term depends on a summation of field gradients and is defined as,

\begin{equation}\label{eqn:Qij}
    {Q}_{ij} = \frac { 1 }{ 2 } \left( \frac { \partial { E }_{ i } }{ \partial j } +\frac { \partial { E }_{ j } }{ \partial i }  \right)
\end{equation}

\noindent and $i, j$ refer to the Cartesian coordinates $x ,y, z$ \cite{han_negative_2009}. There are nine terms in this summation, but only six unique terms, since $Q_{ij} = Q_{ji}$. The MQ field interaction term $G_{ij}$ is defined similarly, replacing electric field gradients with magnetic field gradients. These expressions may be used to determine the power scattered by a spherical particle at any location in an inhomogeneous illuminating field.

Local-field based expressions may also be used to describe higher order multipole interactions; here we restrict discussion to the dipole and quadrupole modes that dominate the response of typical plasmonic and metamaterial NP constituents. Although we focus on scattering cross-sections in this work, expressions for the electromagnetic fields and absorbed power are included in the Supplemental Materials. The local field approach confers a variety of advantages over GLMT. It is particularly useful in cases where the spatial electromagnetic field distributions are already known -- the need for spherical wave or plane wave decompositions is eliminated. Additionally, the local field expressions intuitively reveal opportunities for tuning multipolar light-matter interactions via beam engineering. Below, we demonstrate significant modifications of multipolar scattering spectra and radiation patterns through manipulation of illuminating beam symmetries.

\begin{figure}
\includegraphics[width=8.6 cm]{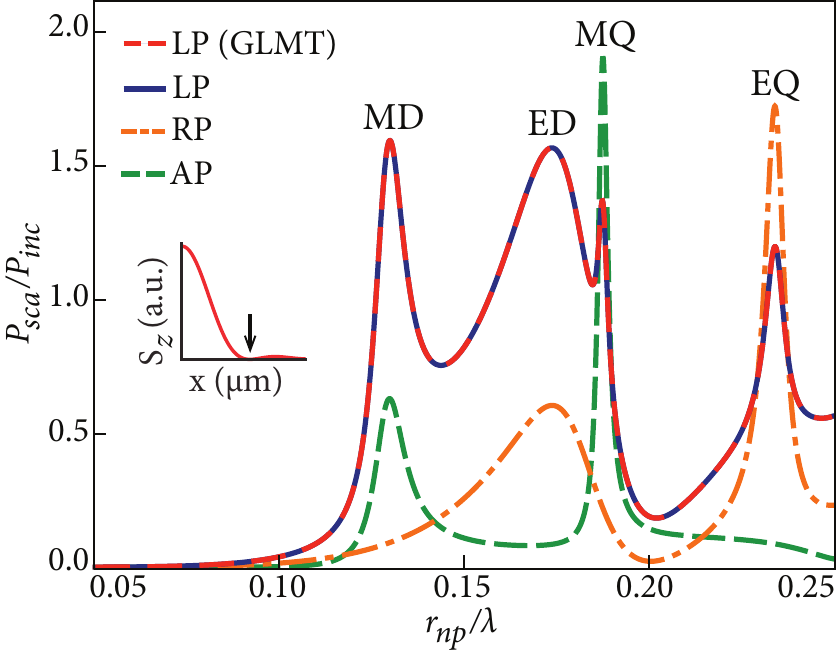}
\caption{\label{fig:figure1}(Color online) Ratio of scattered power to incident power for a spherical NP with refractive index n = 3.7 under linearly (LP, blue line), azimuthally (AP, dashed green line), and radially (RP, dot-dashed orange line) polarized illumination, as a function of normalized frequency. GLMT calculations for scattering by a LP beam (dashed red line) show excellent agreement with the local field expressions derived in this work. Multipolar mode peaks are labeled, with E = electric, M = magnetic, D = dipole, and Q = quadrupole. Inset depicts the normalization scheme in which the incident Poynting vector is integrated over the main intensity lobe of the illumination beam, as indicated by arrow, where Sz is the z-component of the Poynting vector plotted as a function of distance along the x-axis (shown for focused LP illumination).}
\end{figure}

First, the validity of this local field model is demonstrated in Fig. \ref{fig:figure1} for the case of a Silicon NP with index n = 3.7 and radius $r_{NP}$ = 100nm. The ratio of scattered power to incident power is plotted as a function of a normalized frequency parameter: the radius of the NP divided by the wavelength of the incident light. A focused, linearly polarized beam (LP) is expressed as a discrete summation of plane waves. Scattering spectra determined via GLMT (dashed red line) and the local field approach (blue line) are identical. The  ratio of scattered to incident power sometimes exceeds 1 due to the normalization used. Because the incident beam is described as a discrete superposition of plane waves, the incident power transmitted through the xy-plane does not converge (see Supplemental Materials). Here we integrate the incident Poynting vector over a circular area comprising the main intensity lobe of the focused beam, as shown in the inset of Fig. \ref{fig:figure1}.

\begin{table}
\caption{\label{table1}Electric and magnetic field magnitudes and gradients, and orientations of each, under LP, AP, and RP illumination, for ED, MD, EQ, and MQ modes.}
\begin{ruledtabular}
\begin{tabular}{ccccc}
&\multicolumn{2}{c}{Dipoles}&\multicolumn{2}{c}{Quadrupoles}\\
 Illumination & Electric & Magnetic & Electric &
 Magnetic \\
\hline
LP & x & y & $Q_{xz}$ &$Q_{yz}$\\
AP & z & 0 & $Q_{xx}$=$Q_{yy}$ &0 \\
& & & =-2$Q_{zz}$\\
RP & 0 & z & 0 &$Q_{xx}$=$Q_{yy}$\\& & & &=-2$Q_{zz}$ \\
\end{tabular}
\end{ruledtabular}
\end{table}

Illumination by a LP beam results in excitation of all multipole modes, since the relevant dipolar and quadrupolar driving fields are all nonzero (Table \ref{table1}). In contrast, a focused azimuthally polarized (AP) beam exhibits a null in the electric field at the beam's focus \cite{sendur_interaction_2008}. Additionally, although individual electric field gradients are quite strong, the electric quadrupole moment depends on a sum of gradients (equation 4) that cancel each other at the beam's focus. As such, the resultant scattering spectrum (Fig. \ref{fig:figure1}, green line) for a NP at the focus of an AP beam exhibits only magnetic modes. A focused radially polarized (RP) beam exhibits identical symmetries to that of an AP beam, except that the electric and magnetic fields are switched. Consequently, illumination by a RP wave excites only electric modes. The selective excitation of magnetic (electric) modes \cite{mojarad2009tailoring,slaughter2010effects} via azimuthally (radially) polarized illumination is explained here by considering the appropriate local field quantities. By changing symmetries of the illuminating radiation, scattering spectra are strongly modified. 

\begin{figure}
\includegraphics[width=8.6 cm]{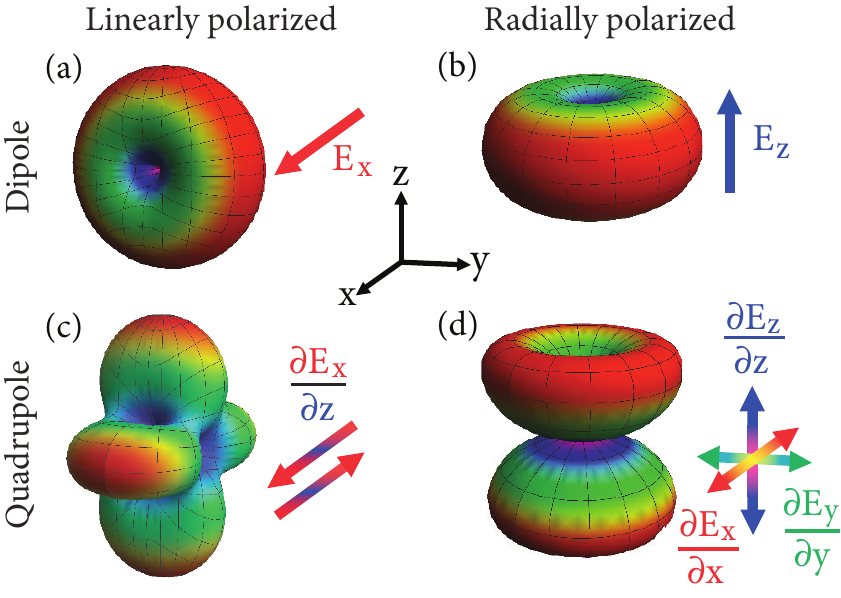}
\caption{(Color online) Radiation patterns of ED (top row) and EQ (bottom row) modes under LP (left column) and RP (right column) illumination, with corresponding field components that drive each excitation. LP illumination results in (a) an x-oriented dipole and RP illumination results in (b) a z-oriented dipole. Due to incident field gradients, LP illumination couples to (c) a transverse quadrupole whereas RP illumination couples to (d) a longitudinal quadrupole mode, which is typically thought to be non-radiating.}
\label{fig:figure2}
\end{figure}

Changing the illumination condition affects not only the scattering spectra of the NP, but also the orientation of the excited multipolar modes. Although the individual multipole resonance frequencies are unchanged, the associated radiation patterns vary with illumination conditions. For example, an x-polarized LP beam excites an in-plane ED, with a dipole moment along the x-axis. The associated radiation pattern is shown in Fig. \ref{fig:figure2}(a). Conversely, the RP wave excites an out-of-plane ED, with a dipole moment along the z-axis, as seen in Fig. \ref{fig:figure2}(b). Thus, changing illumination symmetries allows rotation of the dipole orientation in space. Such modifications of the multipolar radiation patterns are more significant for the case of quadrupole modes, which are driven by field gradients (see Table 1). The x-polarized LP beam only exhibits a non-zero gradient for the x-component of the electric field along the z-direction $\left( \frac { \partial { E }_{ x } }{ \partial z } \neq 0\right)$; $Q_{xz}$ = $Q_{zx}$ are the relevant driving terms. This results in a characteristic four-lobe radiation pattern, as shown in Fig. \ref{fig:figure2}(c). We refer to this as a \textit{transverse} quadrupole. In contrast, the RP beam has only nonzero \textit{longitudinal} gradients: $\frac { \partial { E }_{ x } }{ \partial x } = \frac { \partial { E }_{ y } }{ \partial y } = -2\frac { \partial { E }_{ z } }{ \partial z }$, and $Q_{xx} = Q_{yy} = -2Q_{zz}$ (Table 1). This results in excitation of a longitudinal quadrupole mode. The associated two-lobe radiation pattern (Fig. \ref{fig:figure2}(d)) is markedly distinct from the typical transverse quadrupole mode. Interestingly, this type of quadrupole excitation is typically referred to as a ``dark'' mode in plasmonic dimer antennas \cite{panaro_dark_2014-1,sancho2014boosting}. These results highlight the fact that such modes are only ``dark'' because they cannot be excited by conventional linearly (or circularly) polarized sources. Thus, beam engineering allows for excitation of new classes of multipole modes. 

\begin{figure}[htbp]
\centering\includegraphics[width=8.6 cm]{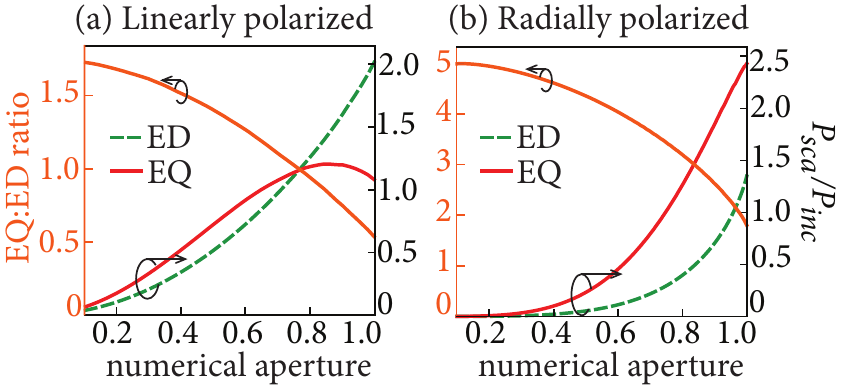}
\caption{(Color online) Change in fraction of scattered power to incident power (right axis) with increasing numerical aperture (i.e., focusing), for ED and EQ modes (on resonance) under (a) LP illumination, and (b) RP illumination. The ED mode generally dominates in LP beams, while the opposite is true of RP beams. The ratio of quadrupole to dipole scattering decreases in both cases (left axis).}
\label{fig:figure3}
\end{figure}

Even when the beam symmetries are unchanged, the multipole spectra depend strongly on other beam properties such as the focused spot size, determined here by changing the numerical aperture (NA = $\sin(\alpha)$). Previously, the focusing angle was fixed to 0.86 ($\alpha = \pm 60^{\circ}$). Varying the NA changes the relative weight of dipole and quadrupole driving terms. This effect is shown in plots of the fraction of incident power scattered by dipole (red) and quadrupole (green) modes, as well as their ratio, under LP (Fig. \ref{fig:figure3}(a)) and RP (Fig. \ref{fig:figure3}(b)) illumination. For the LP beam, the EQ response is generally weaker than the ED response at higher focus, and increases much more slowly with focusing. As a result, the ratio of EQ:ED scattering decreases with tighter focusing. The scaling for RP illumination is quite different: the EQ response is always larger than the ED response (EQ:ED ratio $>$ 1), although the ratio also decreases with tighter focusing. The behavior for the magnetic modes is identical when illuminated by AP beams. Beam engineering not only allows for selective excitation of electric vs. magnetic modes, but also the tuning of the relative weight of quadrupoles vs. dipoles.

\begin{figure}
\centering\includegraphics[width=8.6 cm]{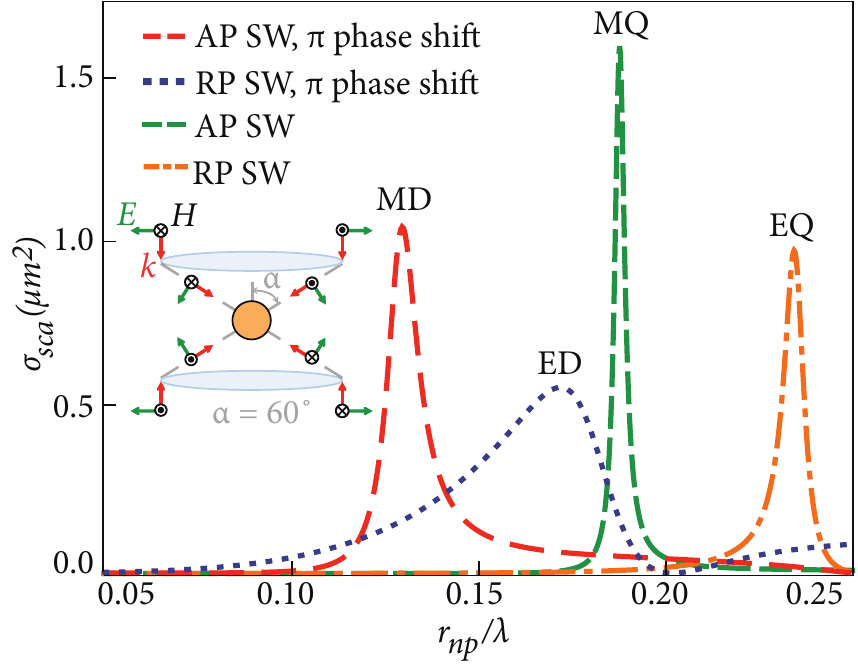}
\caption{(Color online) Scattering cross-section as a function of normalized frequency. Standing wave (SW) illumination by an AP with a $\pi$ phase shift between the two beams (dashed red line), RP with a $\pi$ phase shift (dashed blue line), AP (dashed green line), and RP beam (dashed orange line) beam results in selective excitation of the MD, ED, MQ, and EQ modes, respectively. Inset depicts illumination scheme for RP SW illumination.}
\label{fig:figure4}
\end{figure}

Previous methods for directly quantifying multipolar light-matter interactions have relied on decomposing output light from e.g., photoluminescence \cite{taminiau_quantifying_2012,curto_multipolar_2013} or Rayleigh scattering \cite{wozniak_selective_2015,poutrina_multipole_2014,rolly2013controllable}. Our results suggest an alternative approach: designing light beams that enable selective excitation of individual multipole modes. By taking advantage of the previously discussed beam symmetries we achieve selective excitation of dipole and quadrupole modes in standing wave configurations--i.e., by illuminating from both the top and bottom. Selective excitation of dipole modes requires that all field gradients in the cross terms cancel. This can be achieved by illuminating with two counter-propagating focused RP beams that are $\pi$ out of phase with each other, enabling complete suppression of the magnetic field and both field gradients. A NP placed at the beam focus exhibits only a z-oriented ED mode (see Fig. \ref{fig:figure4}). Similarly, selective excitation of a z-oriented MD mode is achieved via standing wave (SW) illumination comprising of two counter-propagating AP beams that are $\pi$ out of phase with each other. Similarly, selective excitation of the quadrupole modes is achieved by removing the phase difference between two counter-propagating beams. A longitudinal EQ (MQ) mode is excited by SW illumination of two counter-propagating RP (AP) beams. The scattering spectra resulting from these SW illumination profiles is shown in Fig. \ref{fig:figure4}, where the inset depicts the illumination condition for a RP standing-wave without any phase shift. Since these standing waves produce zero power flux (the power from one beam is canceled by the counter-propagating beam) we plot the scattering cross-section: the scattered power divided by the incident intensity. As seen in Fig. \ref{fig:figure4}, individual multipolar resonance spectra are clearly resolved by this illumination method.

In summary, we have derived a simplified method for determining the multipolar scattering of NPs in inhomogeneous beams. Our approach only requires knowing the local electromagnetic fields and their gradients at one point in space, obviating the need for plane wave or spherical wave decompositions. Using this method, we demonstrate the manipulation of multipolar spectra via beam engineering. We show selective excitation of electric and magnetic modes and the emergence of a ``bright'' longitudinal quadrupole. We subsequently demonstrate how to tune the relative weight of dipole and quadrupole contributions by manipulating the beam focusing conditions. We conclude by demonstrating selective excitation of individual multipole modes in standing wave configurations. This work establishes beam engineering as a powerful approach for manipulating the multipolar scattering properties of nanostructures, and for studying the multipolar light-matter interactions in nanoscale elements.

The authors acknowledge support from the Center for Scientific Computing from the CNSI, MRL: an NSF MRSEC (DMR-1121053) and NSF CNS-0960316.

\bibliographystyle{apsrev4-1}

%

\end{document}